\documentclass[journal ]{new-aiaa}
\usepackage[utf8]{inputenc}
\usepackage{textcomp}
\usepackage{natbib}
\usepackage{graphicx}
\usepackage{amsmath}
\usepackage[version=4]{mhchem}
\usepackage{siunitx}
\usepackage{longtable,tabularx}
\title{Satellite Drag Analysis During the May 2024 Gannon Geomagnetic Storm}

\author{William E. Parker\footnote{PhD Candidate, Department of Aeronautics and Astronautics, 77 Massachusetts Avenue, Cambridge, MA 02139} and Richard Linares \footnote{Associate Professor, Department of Aeronautics and Astronautics, 77 Massachusetts Avenue, Cambridge, MA 02139}}
\affil{Massachusetts Institute of Technology, Cambridge, Massachusetts, 02139}

\begin{document}

\maketitle


\begin{abstract}
Between May 10-12, 2024, Earth saw its largest geomagnetic storm in over 20 years. Since the last major storm in 2003, the population of satellites in low Earth orbit has surged following the commercialization of space services and the ongoing establishment of proliferated LEO constellations. In this note, we investigate the various impacts of the geomagnetic storm on satellite operations. A forecast performance assessment of the geomagnetic index $ap$ shows that the magnitude and duration of the storm were poorly predicted, even one day in advance. Total mass density enhancements in the thermosphere are identified by tracking satellite drag decay characteristics. A history of two-line element (TLE) data from the entire NORAD catalog in LEO is used to observe large-scale trends. Better understanding how geomagnetic storms impact satellite operations is critical for maintaining satellite safety and ensuring long-term robust sustainability in LEO. 
\end{abstract}

\section{Introduction}
Between May 7-11, 2024, several X-class solar flares and at least five distinct Earth-bound coronal mass ejections (CMEs) were detected by ground and space-based solar observatories. This increased solar activity originated from AR3664, an active region characterized by a large cluster of sunspots. These events triggered a geomagnetic storm warning from space weather tracking organizations around the world. At around 12:30 PM UTC on May 10, the first of the CMEs reached Earth, leading to a significant geomagnetic enhancement. Widespread auroral activity was reported to be visible during the peak enhancement as far south as 21$^\circ$ N latitude. 

Geomagnetic storms have the potential to cause serious disruptions and failures in safety-critical ground and space-based infrastructure.  Large, unpredictable induced currents along ground-based power transmission lines have historically led to widespread power outages \cite{pulkkinen2005geomagnetic, kappenman2010geomagnetic}. Those same induced currents can also cause sudden satellite electronics failures in orbit \cite{hands2018radiation}. Interruptions in high-frequency radio communication systems are possible \cite{frissell2019high}.  Variability in the structure of the ionosphere during these storms can also lead to uncertain GNSS signal path propagation, affecting the accuracy and reliability of navigation systems  \cite{astafyeva2014geomagnetic}. Increased radiation during the storm puts astronauts in space \cite{cucinotta2014space} and airplane passengers flying near the poles \cite{mertens2010geomagnetic} at risk of dangerously high exposure.

Geomagnetic storms also cause large changes in the structure of the upper atmosphere, both in the charged ionosphere and the neutral thermosphere through the process of Joule heating \cite{sutton2009rapid} (along with particle precipitation \cite{sadler2012auroral} and some other mechanisms). When charged particles from CMEs reach Earth, they interact with the magnetosphere, depositing energy and producing  increased currents in the ionosphere, especially in the auroral regions. As the kinetic energy of the charged particles in the ionosphere increases, they collide more frequently with the neutral particles of the thermosphere. These collisions convert the kinetic energy into thermal energy, which inevitably leads to heating and expansion in the thermosphere. As a result of this Joule heating and additional heating from particle precipitation, the total mass density of the atmosphere at constant altitude can increase by more than an order of magnitude during a geomagnetic storm \cite{forbes2005thermosphere}. Geomagnetic-induced changes in the density of the thermosphere have significant impacts on satellite drag in LEO. Storms can cause forecast errors in satellite positions to reach several kilometers even within a single day \cite{parker2023influences}.

\begin{figure}
    \centering
    \includegraphics[width=0.75\linewidth]{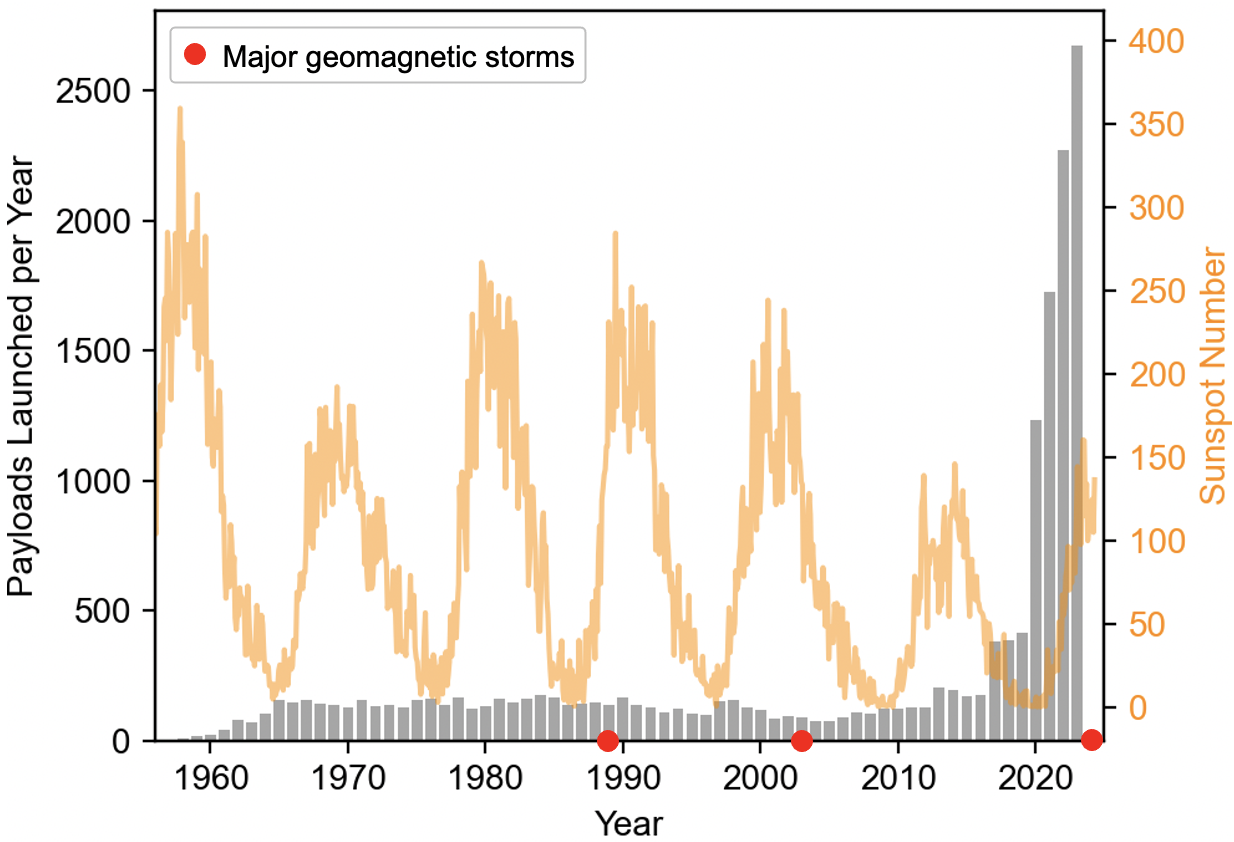}
   \caption{Number of payloads launched to Earth orbit per year. Major geomagnetic storm events, which typically coincide with solar maximum, are marked in red.}
    \label{fig:payload_vs_ssn}
\end{figure}

Figure \ref{fig:payload_vs_ssn} shows a worrying trend in space operations. First, the number of payloads launched into Earth's orbit has grown dramatically since the last major geomagnetic storm (the number of active payloads is somewhat contested, so the annual number of payloads launched, recorded at space-track.org, is a better proxy for the change in activity).  This growth is largely attributable to proliferated LEO constellations like Starlink and OneWeb, but also to the transition towards inexpensive small satellites capable of being launched as ride-shares to orbit. Such a large increase in traffic throughout LEO, along with a ballooning debris population from previous fragmentation events (especially the notable events in 2008 \cite{johnson2008characteristics}, 2009 \cite{kelso2009analysis}, and 2021 \cite{sankaran2022russia}) has made satellite conjunction assessment and collision avoidance a necessary capability on most new spacecraft to ensure satellite safety. 

Geomagnetic storms are most common near solar maximum, the period of peak activity in the roughly 11-year solar cycle. Sun spots, solar flares, and CMEs happen more frequently during solar maximum. Solar extreme ultraviolet (EUV) radiation, the main driver for heating and ionization in the upper atmosphere, also peaks at this time. During major storms in 1989 and 2003, NORAD lost track of many satellites for several days \cite{berger2023thermosphere}. A similar failure today might have dire consequences. Figure \ref{fig:payload_vs_ssn} shows that while there is precedent for geomagnetic storms from 1989 and 2003, the May 2024 Gannon storm is unique in that it is the first to occur during a new paradigm in satellite operations in LEO. As the solar cycle continues to peak throughout 2024 and 2025, continued disruptions to operations are likely to occur. 

It is particularly important that the satellite operator community understands how satellite drag will be impacted during geomagnetic storms as solar maximum approaches. As operators become more dependent on automated collision avoidance systems, it is important to investigate how these systems fare during storms and what the potential consequences might be during prolonged tracking disruptions. In this note, we assess the quality of the geomagnetic activity forecasts leading up to the storm period, then use an empirical model of the upper atmosphere to highlight the structure of density enhancements during the storm. Finally, two-line element (TLE) data from the entire catalog of LEO satellites is used to characterize bulk behavior of satellite operators and debris objects during the storm. Altogether, this work documents the storm's impact on satellite dynamics and may be useful to satellite operators in planning future missions and responding to future storms.

\section{Forecasting Performance}
The dynamics of the upper atmosphere are driven by complex phenomena that are often distilled into a set of scalar activity proxies and indices for the purposes of empirical modeling. F$_{10.7}$ is a useful solar activity proxy measuring the solar radio flux at 10.7 cm \cite{covington1948solar}. It correlates well with solar EUV emissions, which play an important role in driving variability in the mass density of the thermosphere. Another index of particular interest during a geomagnetic storm is the planetary K index, or $Kp$, which is issued once every three hours. $Kp$ is derived from magnetometer measurements at 13 geomagnetic observatories around the world and has been recorded since 1932 \cite{bartels1949standardized}. $Kp$ ranges from 0-9 on a quasi-logarithmic scale, so $ap$, a $Kp$-derived index with a more interpretable linear scaling, is often used in its place. $Kp$ maps directly to $ap$ by a standard conversion \cite{matzka2021geomagnetic}. $ap$ should not be confused with $Ap$, which is the daily average of $ap$. 

\begin{figure}[tb]
    \centering
    \includegraphics[width=0.7\linewidth]{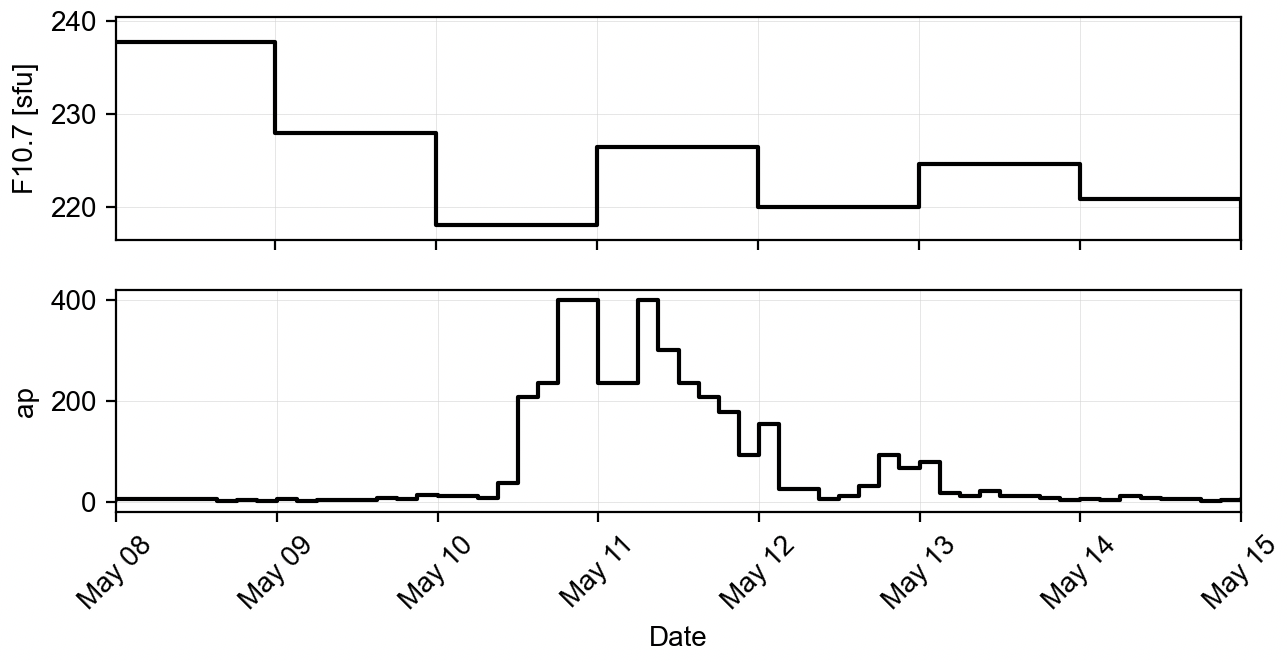}
    \caption{Observed solar and geomagnetic indices during the May 2024 Gannon storm. F$_{10.7}$ is issued daily, while $ap$ is issued every three hours.}
    \label{fig:indices}
\end{figure}

Figure \ref{fig:indices} shows the recorded F$_{10.7}$ and $ap$ during the storm period. F$_{10.7}$ near 220 sfu is very high and leads to an elevated baseline mass density in the thermosphere leading up to the geomagnetic storm. The storm arrives at around midday UTC on May 10, which leads to a large spike in $ap$ to its maximum value of 400. The period of elevated geomagnetic activity lasts nearly two days, then returns to a low-level baseline. 

\begin{figure}[hb!]
    \centering
    \includegraphics[width=0.7\linewidth]{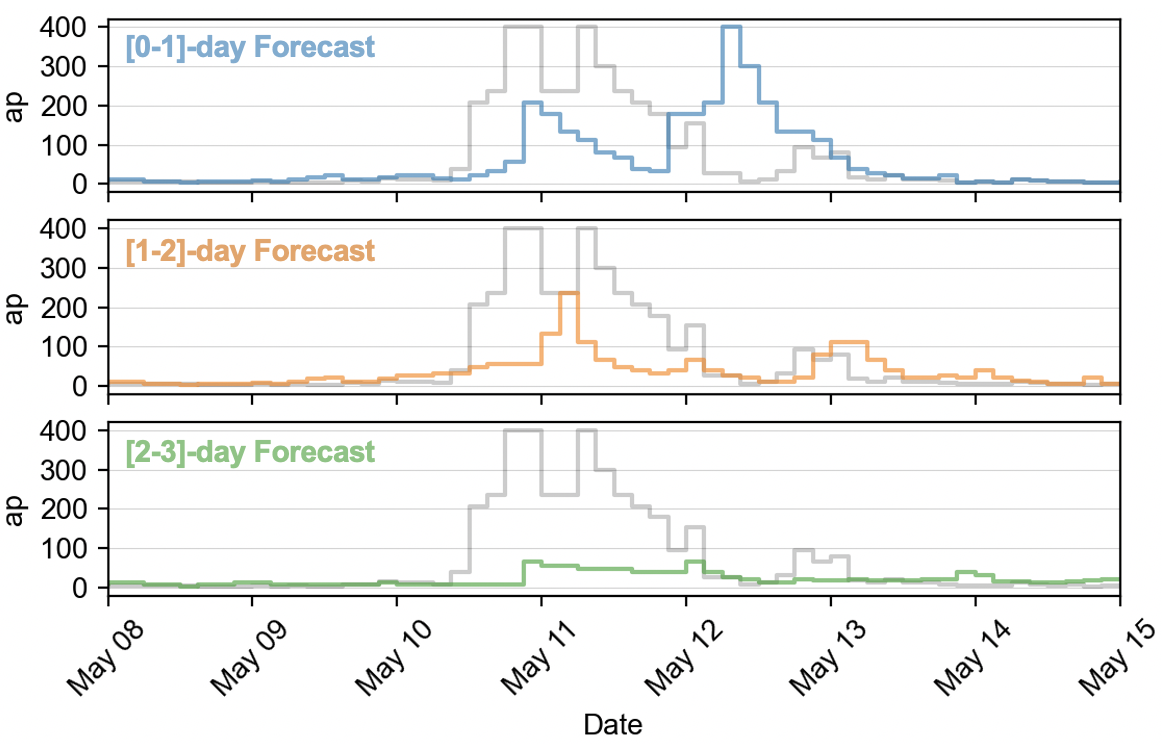}
    \caption{NOAA SWPC's 3-day forecast of $ap$ is issued daily. Forecasts across all lookahead windows struggle, likely due to a strong persistence assumption.}
    \label{fig:forecast}
\end{figure}

Figure \ref{fig:forecast} shows the forecasted $ap$ from May 8 - 15 2024 with the measured $ap$ also shown for reference. NOAA's SWPC releases a 3-day forecast of $ap$ every day. Each forecast includes a predicted value for every 3-hour increment over the next three days. The [0-1]-day forecast represents a forecast that comes from the most recent release, while [1-2]-day represents the forecast release from the day prior, and [2-3]-day comes from the day before that. The $ap$ forecast model works reasonably well during quiet times, but struggled in forecasting this storm. Across all time horizons from 0-3 days, the initial increase in $ap$ was underpredicted by 100-300. After the storm mostly passed on 5/12, the [0-1]-day forecast vastly over-predicted $ap$. These poor forecasts are likely attributable to their dependence on a "persistence" assumption, which considers that the most likely value of the index at some time in the future is close to the current measured value. Most of the best-performing $ap$ index forecast models are mostly based on this persistence principle but also consider other factors including the measured solar wind and observations of the sun's east limb \cite{shprits2019nowcasting}. 

The reality is that forecasting geomagnetic activity is very difficult. Solar energetic particles released in a coronal mass ejection (CME) move at speeds between $250-3000$ km/s \cite{paouris2021assessing}. Some CMEs reach Earth as quickly as 15 hours after ejection, but most take closer to three days. Earth-based and space-based telescopes can observe CMEs and take measurements of the solar wind to help forecast geomagnetic enhancements, but those forecasts only have the potential to be useful in the short-term since longer-term forecasting would mean predicting the CMEs themselves. 

Uncertainty in propagated satellite state may arise from several potential sources. These sources include initial measurement error, space weather index forecasting error, atmospheric density modeling error, and satellite ballistic coefficient error. In this case, since the forecast for $ap$ is so poor, it would not be surprising if the space weather index forecasting error is the dominant term in propagated satellite state uncertainty. Being able to accurately propagate a spacecraft, or at least bound the set of future states with a reasonable forecast uncertainty, is a critical component for performing useful conjunction assessment in LEO. 

\section{Modeled Density Enhancements}

\begin{figure}[h!]
    \centering
    \includegraphics[width=1\linewidth]{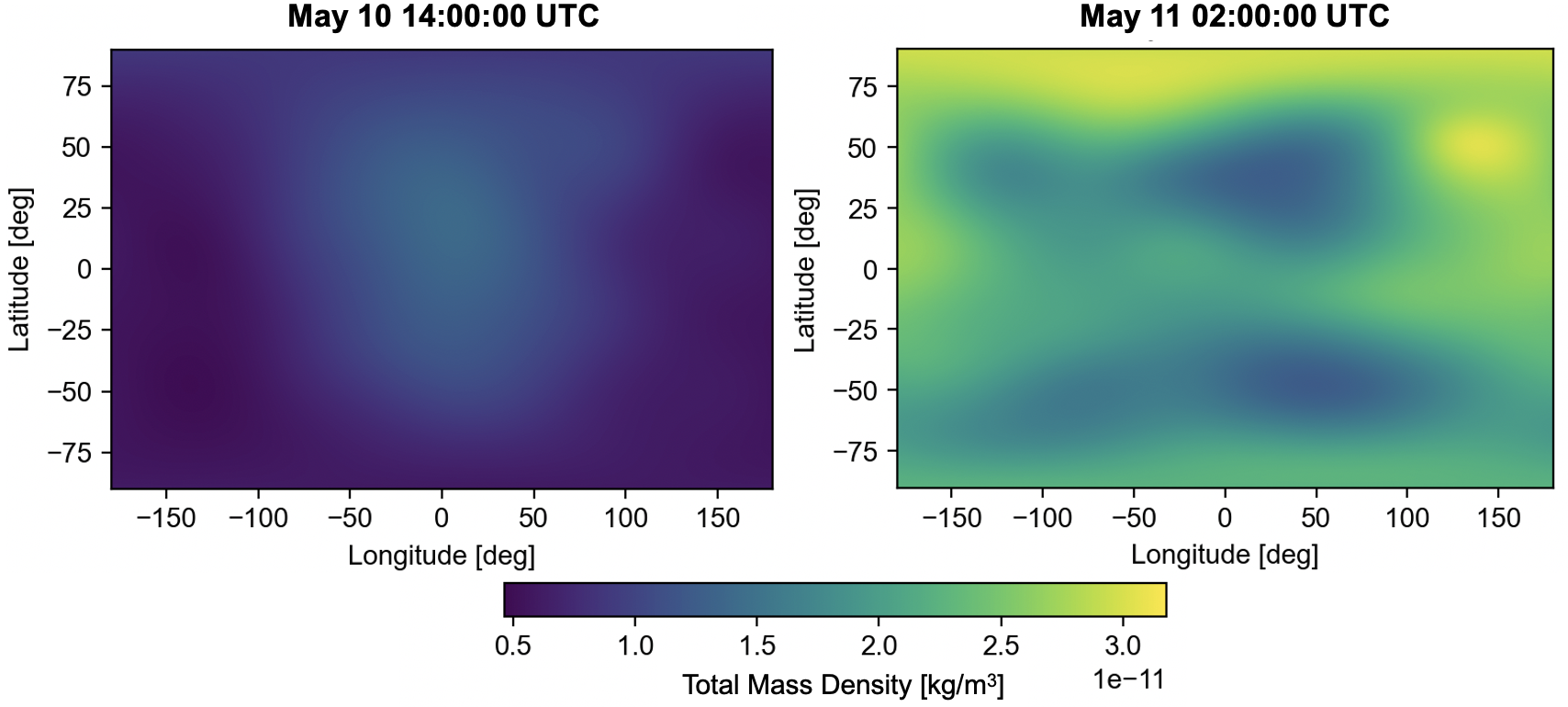}
    \caption{Total mass density at 400 km altitude before and during the storm from NRLMSISE-00. }
    \label{fig:dens_compare}
\end{figure}

One of the only viable approaches for estimating the mass density of the thermosphere is through in-situ measurements from spacecraft in LEO. Satellites with onboard accelerometers like CHAMP \cite{reigber1999champ}, GRACE \cite{davis2000grace}, and Swarm \cite{doornbos2009air} infer satellite drag from accelerations measured along their respective trajectories. Swarm A, B, and C are also fitted with a GNSS receiver and publish a history of measured satellite states at a rate of 1 Hz, which has also been used to infer satellite drag \cite{gondelach2021real}. In practice, we rely heavily on models of the upper atmosphere to predict satellite motion through complex density fields during geomagnetic storms. The best of these models are physics-based, but they struggle with long run times and are therefore difficult to implement in practice for many satellite operators. Empirical models, by comparison, are sometimes less accurate but are fast to evaluate. One such empirical model is the US Naval Research Laboratory's Mass Spectrometer and Incoherent Scatter radar (Extended) model, or NRLMSISE-00 \cite{picone2002nrlmsise}, which takes F$_{10.7}$ and $ap$ as inputs to compute thermosphere properties, including the total mass density. Figure \ref{fig:dens_compare} shows the NRLMSISE-00 derived total mass density at 400 km on May 10 14:00:00 UTC and 12 hours later on May 11 02:00:00 UTC. Before the storm hits, only slight density enhancements from diurnal heating of the atmosphere are apparent. Once the storm arrives, Joule heating and particle precipitation create large density enhancements of up to 6x the baseline value 12 hours prior. Most of the density enhancement is focused in the northern hemisphere. The accuracy of NRLMSISE-00 is limited by its simplicity, only considering two main drivers. Still, the rough estimate for the density increases seem appropriate given observations of enhanced satellite drag decay during the storm. 

\section{Operational Implications}
\begin{figure}[b!]
    \centering
    \includegraphics[width=0.65\linewidth]{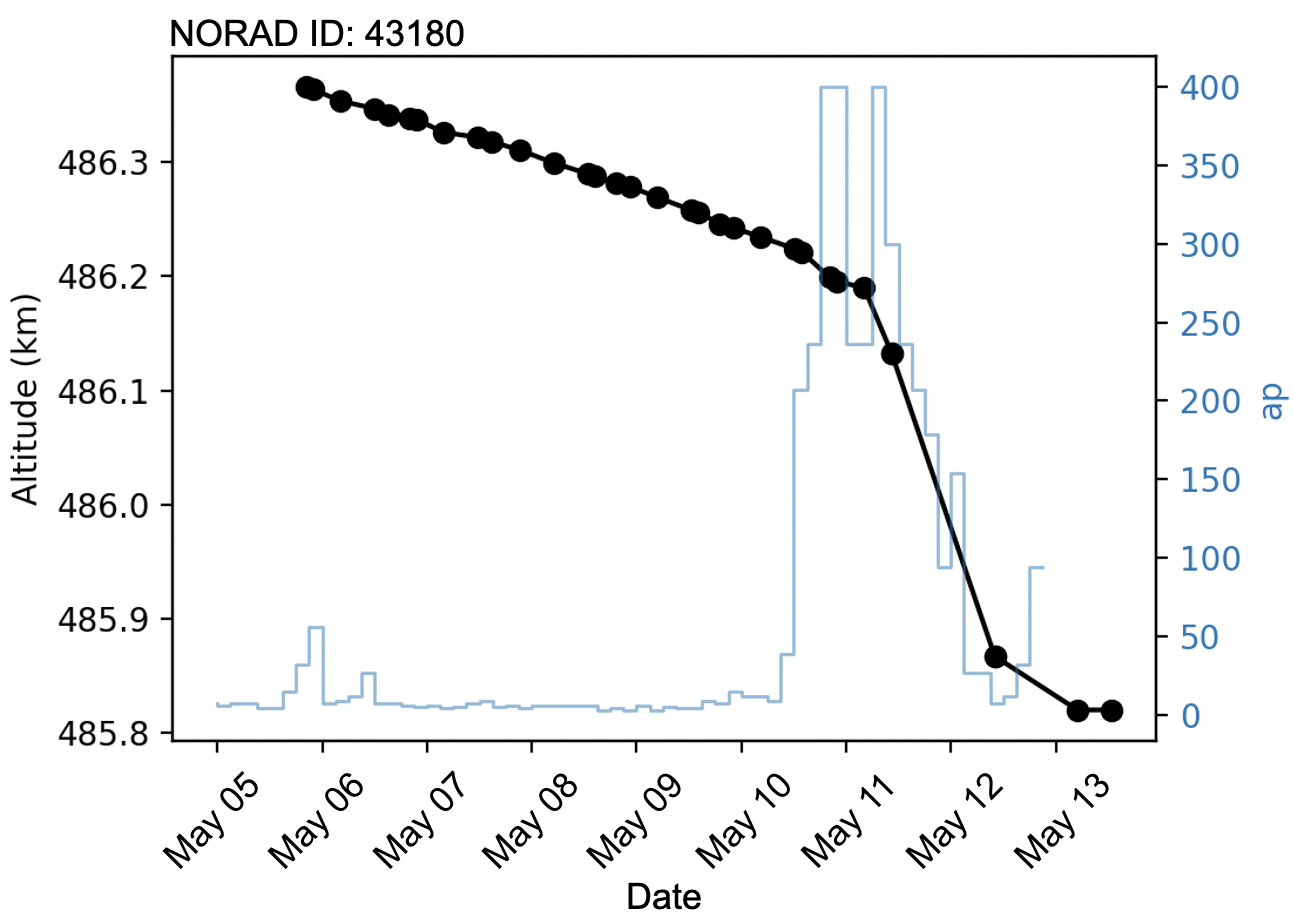}
    \caption{Time-averaged orbit altitude for SATCAT 43180, or KANOPUS-V 3, from TLEs leading up to and during the storm.}
    \label{fig:tle_decay}
\end{figure}

Most tracked objects in LEO showed some signs of increased orbital decay during the period of geomagnetic enhancement. Figure \ref{fig:tle_decay} shows the the time-averaged orbit altitude of SATCAT 43180 (KANOPUS-V 3) from TLEs before, during, and after the storm. Before the storm, the object passively decayed at a rate of approximately 38 m/day. During the storm, however, the decay rate increased more than 4x to 180 m/day. The cadence of TLE publishing during the storm dropped while the object was undergoing the period of rapid decay. 

\begin{figure}
    \centering
    \includegraphics[width=\linewidth]{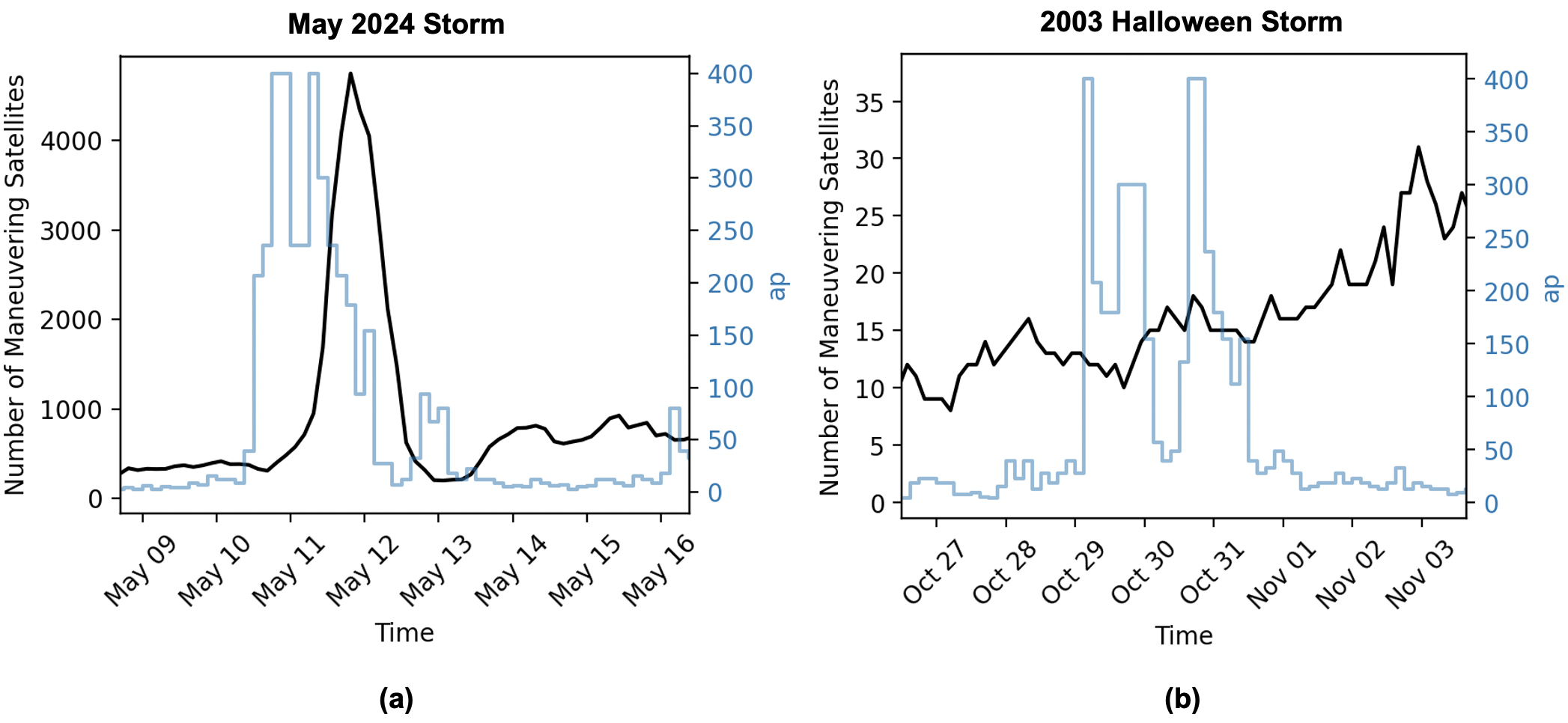}
    \caption{(a) Satellites in LEO (mostly SpaceX) maneuver \emph{en masse} in response to the sudden loss in altitude during the May 2024 Gannon storm. (b) During the 2003 Halloween storm, there was no comparable change in satellite maneuver activity (note the difference in scaling). }
    \label{fig:Maneuver}
\end{figure}

For many operators managing satellites during the storm, such a sudden drop in orbital altitude is untenable. Unplanned orbital decay can disrupt constellations by causing uneven satellite altitudes, which results in undesirable orbit phasing in the short term and relative plane drift over the long term. Other satellites performing Earth observation tasks may also have similarly tight constraints on orbital altitude and require regular station-keeping. Figure \ref{fig:Maneuver} shows the number of LEO tracked objects maneuvering over time with the time history of $ap$ in the background for reference. The figure includes both the May 2023 storm and the October 2003 Halloween storm. A maneuver is counted when the mean orbit altitude of an object increases by > 100 m over a 3-hour window. For the May 2023 storm, about 300 of the nearly 10,000 active payloads in LEO appear to be maneuvering during the quiet period leading up to the storm. After the storm hits, with some offset to account for the time it takes for drag decay to accumulate, thousands of satellites begin to maneuver \emph{en masse} in response to the sudden increase in atmospheric density. For comparison, maneuver activity in LEO was not comparably impacted during the October 2003 Halloween storm. Most of the May 2024 maneuver activity is attributable to the Starlink constellation, which performs autonomous orbit maintenance and thus responds quickly to perturbing events. Onboard orbit maintenance will become more common as other proliferated LEO constellations are established. 

The satellite conjunction assessment process typically starts by considering a look-ahead window of seven days and propagating every tracked object forward to screen for potential conjunctions. It is good practice for satellite operators to provide tracking agencies with ephemeris files that include planned station-keeping or collision avoidance maneuvers during the look-ahead window that may impact future satellite states. The station-keeping maneuvers that occurred following the storm were certainly not planned more than a few hours in advance since forecasts of the storm were poor even a day prior to the event. Many potential conjunctions that were anticipated before the storm were likely impacted by this \emph{en masse} maneuver since most tracked satellites would have ended up in very different positions at the time of the conjunction. After so many satellites maneuver at once, the conjunction assessment pipeline needs to start over from new initial satellite states after the group maneuver and after the storm has passed. These challenges in handling both the poor drag forecasts and the unplanned station-keeping maneuvers call into question the capabilities of the existing conjunction assessment procedures during geomagnetic storm conditions. As we become more dependent on this infrastructure to maintain safety in LEO, it needs to be made more robust to these geomagnetic storms. 

\section{Geomagnetic Storms as Debris Sinks}
\begin{figure}[h!]
    \centering
    \includegraphics[width=0.8\linewidth]{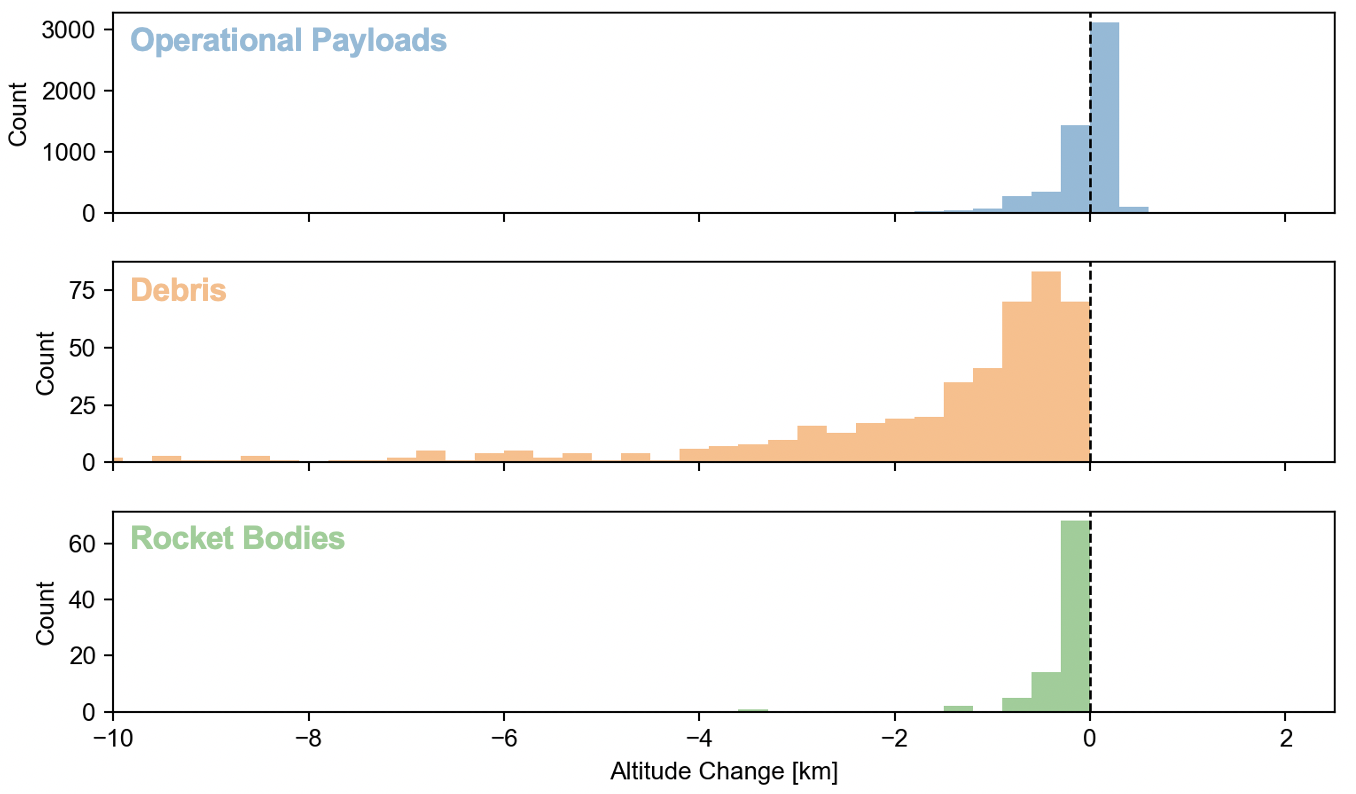}
    \caption{Altitude change for cataloged objects by type within 400-700 km between 10 May and 13 May 00 GMT.}
    \label{fig:decay_by_type}
\end{figure}

While storms represent a challenge for performing actionable conjunction assessment, they also offer a unique benefit to the operating environment in LEO. The increase in thermospheric total mass density during the storm leads to enhanced orbital decay across most tracked objects in the satellite catalog (especially those at lower altitudes). Only active satellites, however, are capable of performing orbit-maintenance maneuvers. Figure \ref{fig:decay_by_type} shows the distribution of altitude change for satellites within 400-700 km altitude between 5/10 and 5/13 at 00 GMT. In general, maneuverable operational payloads maintain their altitude by performing orbit-raising maneuvers in the wake of the storm. However, debris objects and rocket bodies both see a period of substantial altitude decay. Debris objects generally decay the fastest because the population has the highest average $A/m$ of the three groups considered.  

A positive insight from this storm is that it helped to hasten the decay of debris objects from orbit while most satellites escaped relatively unaffected. Debris is notoriously difficult to remove, so a strong solar cycle with strong geomagnetic storms is one of the best things for helping to maintain a long-term operable environment in LEO. 

\section{Conclusion}
This note highlighted impacts from the May 2024 Gannon geomagnetic storm on satellite operations in LEO. The storm represented a serious challenge for the existing conjunction assessment infrastructure as it produced large, unpredictable perturbations on satellite trajectories in LEO.  New proliferated LEO constellations require tight station-keeping bounds to prevent undesirable orbit phasing. Automated station-keeping, especially from the Starlink constellation, caused nearly half of all the active satellites in LEO to maneuver at once in response to the storm. The combination of unpredictable satellite drag and bulk maneuvering made it very difficult or impossible to identify potential conjunctions during the storm and in the days that followed. While the storm represented a risk to the LEO environment in the short term, it also helped to hasten the removal of debris populations from orbit. This passive debris removal is critical for the long-term sustainability of operations in LEO. Moving forward, it is important that we recognize the limits that the environment imposes on satellite activity in LEO. Operators and regulators should consider the robustness of the conjunction assessment infrastructure to events of this kind when deciding how much to rely on it.

\section{Acknowledgments}
This work was supported by the National Science Foundation Graduate Research Fellowship under Grant No. 1745302 and the joint NSF-NASA Space Weather with
Quantified Uncertainties program under NSF grant number 2028125. The authors gratefully acknowledge the sponsors for their support.

\newpage
\bibliography{main}

\end{document}